\documentclass[prl,twocolumn,showpacs,amsmath,superscriptaddress,psfig,amssymb]{revtex4}
\usepackage{amssymb}

\usepackage{graphicx}
\usepackage{dcolumn}
\usepackage{bm}

\begin{document}


\title{Multiple quantum phase transitions in a heavy fermion antiferromagnet}

\author{L. Jiao}
\affiliation{Center for Correlated Matter and Department of Physics, Zhejiang University, Hangzhou,
Zhejiang 310027, China}
\author{H. Q. Yuan}
\email{hqyuan@zju.edu.cn}
\affiliation{Center for Correlated Matter and Department of Physics,
Zhejiang University, Hangzhou, Zhejiang 310027, China}
\author{Y. Kohama}
\affiliation{Los Alamos National Laboratory, Los Alamos, NM 87545}
\author{E. D. Bauer}
\affiliation{Los Alamos National Laboratory, Los Alamos, NM 87545}
\author{J. -X. Zhu}
\affiliation{Los Alamos National Laboratory, Los Alamos, NM 87545}
\author{J. Singleton}
\affiliation{Los Alamos National Laboratory, Los Alamos, NM 87545}
\author{T. Shang}
\affiliation{Center for Correlated Matter and Department of Physics, Zhejiang University, Hangzhou,
Zhejiang 310027, China}
\author{J. L. Zhang}
\affiliation{Center for Correlated Matter and Department of Physics, Zhejiang University, Hangzhou,
Zhejiang 310027, China}
\author{Y. Chen}
\affiliation{Center for Correlated Matter and Department of Physics, Zhejiang University, Hangzhou,
Zhejiang 310027, China}
\author{H. O. Lee}
\affiliation{Los Alamos National Laboratory, Los Alamos, NM 87545}
\author{T. Park}
\affiliation{Department of Physics, Sungkyunkwan University, Suwon 440-746, Korea}
\author{M. Jaime}
\affiliation{Los Alamos National Laboratory, Los Alamos, NM 87545}
\author{J. D. Thompson}
\affiliation{Los Alamos National Laboratory, Los Alamos, NM 87545}
\author{F. Steglich }
\affiliation{Max-Planck-Institut f$\ddot{u}$r Chemische Physik fester Stoffe,
N$\ddot{o}$thnitzer Str. 40, 01187 Dresden, Germany}
\author{Q. Si}
\email{qmsi@rice.edu}
\affiliation{Department of Physics and Astronomy, Rice University, Houston, TX 77005}

\date{\today}

\begin{abstract}

We report measurements of magnetic quantum oscillations and specific heat at low temperatures across a field-induced antiferromagnetic quantum critical point (QCP) ($B_{c0}$$\approx$50T) of the heavy-fermion metal CeRhIn$_5$. A sharp magnetic-field induced Fermi surface reconstruction is observed inside the antiferromagnetic phase. Our results demonstrate multiple classes of QCPs in the field-pressure phase diagram of this heavy-fermion metal, pointing to a universal description of QCPs. They also suggest that robust superconductivity is promoted by unconventional quantum criticality of a fluctuating Fermi surface.
\end{abstract}


\maketitle

Conventional, thermally-driven continuous phase transitions are described by universal critical behavior that is independent of microscopic details of a specific material. Current studies on a growing set of materials have focused on quantum-driven phase transitions that occur at absolute zero temperature. Whether universality also applies to continuous quantum phase transitions, i.e., quantum critical points (QCPs), and how to properly characterize them are important open issues\cite{1Sachdev,2Stewart,3von,4Si}. It was proposed recently that Fermi surface reconstruction can be used to distinguish different types of QCPs in heavy fermion compounds\cite{5Si,6Coleman}. However, direct experimental evidence for this proposal remains to be established.

The intermetallic heavy-fermion metals, whose ground states can be tuned readily by a non-thermal control parameter, such as magnetic field, pressure or chemical composition, represent a prototype for studying QCPs\cite{2Stewart,3von}. Despite examples of a QCP in many different heavy-fermion systems, a definitive theory, analogous to that for thermally-driven phase transitions, has not emerged, though two distinctly different sets of models of a QCP have been proposed. A distinguishing characteristic of these models is the response of electronic degrees of freedom. An extension of the theory of thermally-driven transitions to the zero-temperature limit of a continuous QCP considers only fluctuations of an order parameter, for example, the sublattice magnetization of a spin-density wave (SDW), as critical excitations\cite{8Hertz,9Moriya,10Millis}. Electronic degrees of freedom do not become critical when the SDW transition is tuned to zero temperature and, consequently, the Fermi surface smoothly evolves across the QCP. Though there is experimental support for this idealized model of a QCP, e.g., CeCu$_2$Si$_2$\cite{11Arndt}, the quantum critical response of other heavy-fermion materials is clearly inconsistent with its predictions and questions the universality of its simplifying assumptions. A qualitatively different model predicts a sharp reconstruction of the Fermi surface while crossing the QCP\cite{12Si,13Coleman,14Senthil} due to the essential involvement of the electronic degrees of freedom. Such an unconventional QCP has been proposed to involve the critical destruction of the Kondo effect, in addition to the fluctuations of the order parameter\cite{4Si}. Evidence for this type of QCP has been found in three heavy-fermion systems UCu$_{5-x}$Pd$_x$\cite{15Aronson}, CeCu$_{6-x}$Au$_x$\cite{16Schroder}, and YbRh$_2$Si$_2$\cite{17Paschen}. In each case, however, direct evidence for a change in Fermi surface structure is lacking. Besides the need to verify the basic prediction of this alternative model of QCPs, a further test of its validity would be the more restrictive observation of a change in the Fermi surface as a function of multiple tuning parameters.

\begin{figure}[b]\centering
 \includegraphics[width=8.5cm]{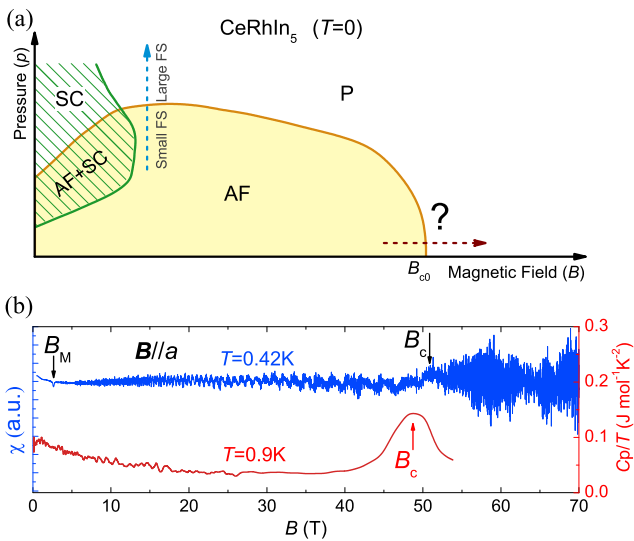}
\caption{(Color online) (a), Schematic magnetic field-pressure phase diagram of CeRhIn$_5$ at zero temperature. Pressure suppresses the AF order and induces superconductivity (SC), leading to several ground states (i.e., AF order, SC and their coexistence) in the phase diagram\cite{19Park,20Knebel}. A pressure-induced change from a small to a large Fermi surface and a seemingly diverging effective mass were observed at the AF QCP for fields larger than the superconducting upper critical field \cite{7Shishido}. (b), Specific heat and dHvA oscillations in a pulsed magnetic field. The magnetic susceptibility ($T$=0.42K) and ac specific heat ($T$=0.9K) of CeRhIn$_5$ are shown as a function of magnetic field up to 72T and 55T, respectively. They display a metamagnetic transition at $B_M$$\approx$2.5T, and a transition from the AF phase to the paramagnetic (P) one at a higher field $B_c(T)$.}
\label{Fig.1}
\end{figure}

CeRhIn$_5$, a heavy-fermion antiferromagnet with a N$\acute{\textrm{e}}$el temperature $T_N$$\approx$3.8K at ambient pressure\cite{18Hegger}, is ideally suited for these purposes. As shown in the schematic phase diagram at zero temperature [Fig. 1(a)], application of pressure suppresses the antiferromagnetic (AF) order and induces superconductivity over a wide pressure region\cite{19Park,20Knebel}. In the presence of a modest magnetic field sufficiently large to suppress superconductivity, an AF QCP is exposed through pressure tuning\cite{19Park,20Knebel}. A sharp change of the Fermi surface is observed across this pressure-induced QCP via de Haas-van Alphen (dHvA) oscillations [Fig. 1(a), blue arrow]\cite{7Shishido}. Because the AF order at ambient pressure is robust against magnetic field\cite{21Shishido}, CeRhIn$_5$ allows for measurements of magnetic quantum oscillations across its field-tuned QCP as well, thereby providing a rare system in which the nature of QCPs can be probed under multiple tuning parameters.

In this Letter, we have characterized the magnetic quantum phase transition of CeRhIn$_5$ at ambient pressure by measuring the ac specific heat and dHvA oscillations in a pulsed magnetic field up to 72T as shown in Fig. 1(b) [see Supplementary Material (SM)]. These measurements allow explicit mapping of the magnetic field-temperature phase diagram and the investigation of the evolution of the Fermi surface as a function of magnetic field. A sharp Fermi surface reconstruction is observed in the AF state while approaching an AF QCP at $B_{c0}$$\approx$50T.

Figure 2 shows the three-dimensional (3D) plot of the specific heat $C_p/T$ as a function of magnetic field and temperature for CeRhIn$_5$ (\textbf{B}$\parallel a$). The pronounced maximum in $C_p(B)/T$ marks the onset of the N$\acute{\textrm{e}}$el transition. The magnetic phase boundary $T_N(B)$ of CeRhIn$_5$ is then derived by projecting the peak positions of $C_p(B)/T$ onto the \textit{B-T} plane. This, along with the field-dependence of the magnetic susceptibility as illustrated in Fig. 1(b), yields a zero-temperature critical field $B_{c0}=B_c$ ($T\rightarrow$0) of about 50 T.

\begin{figure}[b]\centering
 \includegraphics[width=8.5cm]{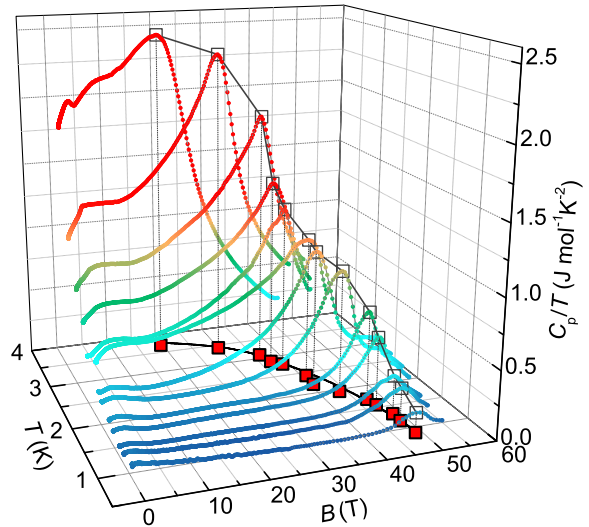}
\caption{(Color online) Temperature and magnetic-field dependence of the specific heat $C_p/T$ for CeRhIn$_5$. The N$\acute{\textrm{e}}$el temperatures, $T_N(B)$, are determined from the peak values of $C_p(B)/T$, which denote a lower bound of the magnetic phase transition. The weak kinks at $B_M$$\approx$2.5T are attributed to the metamagnetic transition whose temperature is consistent with that derived from the dHvA measurements, as shown in Fig.1(b). The small hump at $B$=5-10T, which becomes less pronounced upon cooling, is attributed to residual addenda contributions.}
\label{Fig.2}
\end{figure}

We now analyze the field-dependent evolution of the dHvA frequencies, which are determined by the extremal orbits of the Fermi surface. In Figs. 3(a)-3(d), we show the fast Fourier transform (FFT) spectra of dHvA oscillations obtained from the magnetic field windows of 10T$<B<$40T [(a) and (b)] and 50T$<B<$70T [(c) and (d)] at $T$=0.5K ($B_c$$\approx$49T) and $T$=1K ($B_c$$\approx$47T), respectively. In the case of 10T$<B<$40T and $T$=0.5K, peaks, labeled $f_1$, $f_2$, $f_3$ and $f_4$, in the dHvA spectrum are located in the frequency region 200T-1000T, which agree well with previous results\cite{22Cornelius} and are characteristic of a ¡®small¡¯ Fermi surface that does not include the 4$f$-electron of Ce. Several harmonic frequencies are observed for the most pronounced dHvA peak at $f_4$$\approx$790T. According to the Lifshitz-Kosevich formula \cite{LK}, increasing temperature causes a strong damping of the oscillation amplitude. Indeed, at the higher temperature, $T$=1 K, we only observe the dominant peak $f_4$ [Fig. 3(b)] as well as $f_2$ and the harmonic frequencies of $f_4$ [Fig. 3(d)]. The precise determination of the cyclotron mass $m^{\ast}$ will require future measurements at additional temperatures. Strikingly, we observe four new dHvA branches ($f_5$, $f_6$, $f_7$ and $f_8$) with much larger frequencies in the FFT spectra over the magnetic field range of 50T-70T at $T$=0.5K [Fig. 3(c)], indicating that the volume of electronic states in $k$-space enclosed by the Fermi surface has undergone a pronounced increase upon the application of a magnetic field. A similar reconstruction of the Fermi surface is still observed at $T$=0.8K, but cannot be resolved anymore around $T$=1K (cf. SM).

\begin{figure}[b]\centering
 \includegraphics[width=8.5cm]{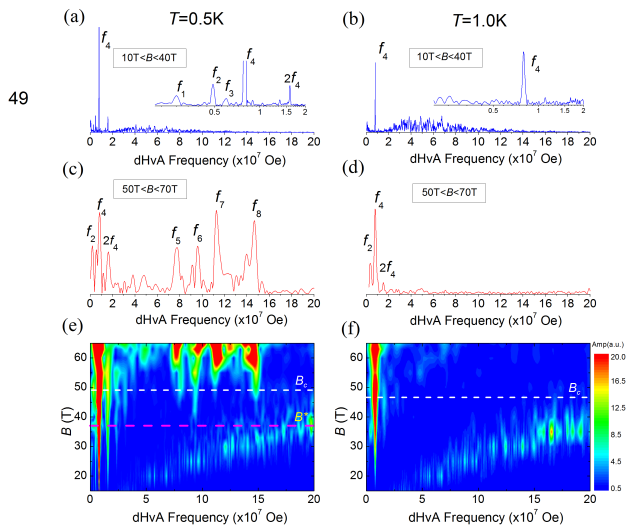}
\caption{(Color online) Fast Fourier transform (FFT) spectra of the dHvA oscillations for CeRhIn$_5$ at $T$=0.5K (left panel) and $T$=1K (right panel) (\textbf{B}$\parallel a$). (a)-(d): FFT spectra determined in a field range of 10T$<\textit{B}<4$0T [(a) and (b): \textit{B}$<B_c(T)$] and of 50T$<\textit{B}<$70T [c and d: $B>B_c(T)$], respectively. In the latter field range, new dHvA oscillations with much larger frequencies ($f_5$, $f_6$, $f_7$ and $f_8$) compared to those of the smaller-field range are observed at $T$=0.5K but are absent at $T$=1K. (e)-(f): Detailed evolution of the dHvA frequencies with magnetic field. The FFT was performed in a field interval of $\Delta$$B$=10T centered around a given field point. The color represents the dHvA amplitude, with the maximum shown in red and the minimum in blue. The broad shadow which shifts to higher frequencies with increasing magnetic field is attributed to the background contributions. Dashed lines indicate the AF critical field $B_c$ (white) and the onset field $B^{\ast}$ of the large frequencies (pink) at the corresponding temperatures.}
\label{Fig.3}
\end{figure}

In order to probe more precisely the onsets of the field-induced new dHvA branches, we examine the dHvA frequencies as a function of magnetic field over finer field steps. Figure 3(e) ($T$=0.5K) and Fig. 3(f) ($T$=1K), where color illustrates the dHvA amplitudes, reveal several important features. First, low dHvA frequencies with sharp peaks exist in the low magnetic field region at $T$=0.5K. As the field is increased, their amplitudes grow, but their widths broaden significantly due to the loss of spectral resolution. Peaks of similarly low frequencies seem to exist up to the highest magnetic field measured, even though they become increasingly difficult to resolve in the FFT spectra. Second, the new dHvA peaks at large frequencies set in at a field of $B^{\ast}$(0.5K)$\approx$37T, which is well below the critical field $B_c$(0.5K)$\approx$49T for the AF order. These dHvA frequencies vary from about 8,000T to 15,000T, well beyond those observed at low fields. Third, at $T$=1K, the large-frequency peaks are no longer observable; only the dominant small-frequency branches, e.g., $f_4$=790T and its harmonics, remain visible.

Figure 4(a) presents the temperature-magnetic field phase diagram for CeRhIn$_5$. The triangles mark the metamagnetic transition around 2.5T and the square symbols show the field dependence of the N$\acute{\textrm{e}}$el temperature, $T_N(B)$. The latter was determined from both the specific heat measured in a Quantum Design Physical Properties Measurement System (open squares) and in a pulsed field (red squares), and the dHvA oscillations (green squares). Upon applying a magnetic field, the AF transition at $T_N$ first exhibits a slight increase, but continuously decreases for fields larger than about 10T down to the lowest accessible temperature of $T$$\approx$0.4K. These observations provide strong evidence for an AF QCP around $B_{c0}$$\approx$50T. Note that the data points near $B_{c0}$ follow the expression of $T_N$$\sim$($B_{c0}-B$)$^{2/3}$ (see the dotted line), expected for a 3D-SDW QCP \cite{1Sachdev}. Remarkably, the system undergoes a sharp change of the Fermi surface at $B^{\ast}(T)$ inside the AF phase. The hatched area above $T^{\ast}(B)$ [=$T(B^{\ast})$] denotes the uncertainties of the crossover boundary as a result of the limited number of temperature points we could measure and also the possible self-heating effects. For $B<B^{\ast}$, previous dHvA studies have shown that the Ce-4$f$ electrons are localized in CeRhIn$_5$ and do not contribute to the Fermi sea\cite{21Shishido,23Harrison}. Our results are compatible with this conclusion. Above $B^{\ast}(T)$, our dHvA oscillations occur not only at similarly low frequencies but also at larger frequencies, about 10$^4$T. The latter are comparable with those measured in CeCoIn$_5$\cite{24Hall} and our calculations of CeRhIn$_5$ assuming itinerant 4$f$-electrons (cf. SM). This suggests that the Fermi surface change across $B^{\ast}(T)$ is not due to a change in the magnetic structure, but instead is caused by the delocalization of 4$f$-electrons in CeRhIn$_5$. Thus, the temperature scale $T^{\ast}(B)$ [=$T(B^{\ast}$)] at $B>B_0^{\ast}(T=0)$ separates regimes of large and small Fermi surfaces. This is also consistent with our observations that raising temperature at large fields quickly suppresses the dHvA peaks at large frequencies; the small value for $T^{\ast}$ implies a large quasiparticle mass, providing further evidence for the direct involvement of the 4$f$-electrons in the corresponding branches of Fermi surface. At low temperatures, the persistence of the large Fermi surface up to, at least, $B$=72T can be understood by a simple consideration of the magnetic field and temperature scales. The temperature for the onset of Kondo screening in CeRhIn$_5$ is about 10K\cite{18Hegger}, more than twice the N$\acute{\textrm{e}}$el temperature. Given that the critical magnetic field $B_{c0}$ is about 50T, the single-ion Kondo field is expected to be substantially larger than the presently accessible fields.

\begin{figure}[b]\centering
 \includegraphics[width=8cm]{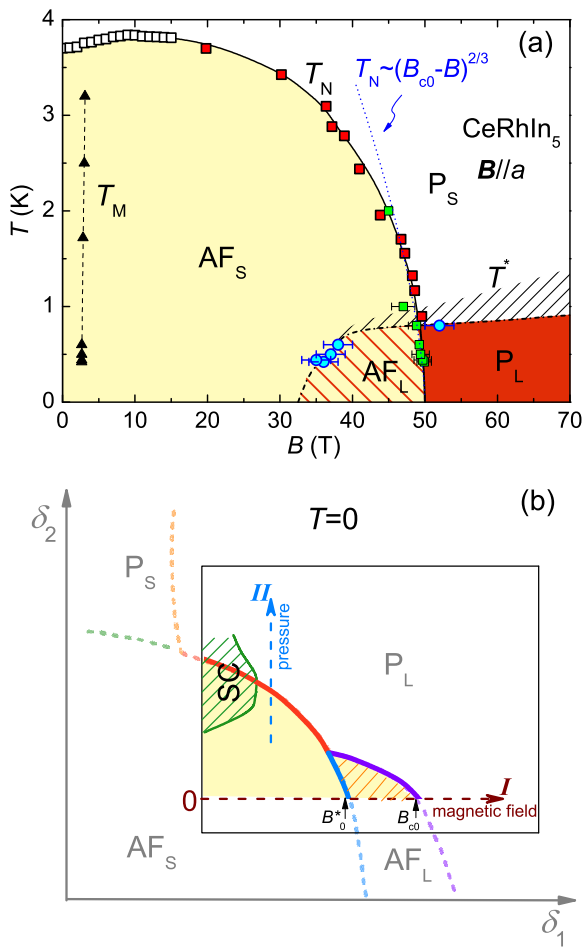}
\caption{(Color online) (a): Temperature-magnetic field phase diagram of CeRhIn$_5$ at ambient pressure. The phases AF$_\textrm{S}$, AF$_\textrm{L}$, P$_\textrm{S}$ and P$_\textrm{L}$ contain two kinds of distinctions: antiferromagnetic (AF) or paramagnetic (P) state on the one hand, and Kondo screening (large FS, ¡°L¡±) or Kondo destruction (small FS, ¡°S¡±) on the other hand. The error bars of $B^{\ast}(T)$ are determined by the magnetic field intervals, $\Delta B =5$T, used in our FFT analysis. (b): Placing CeRhIn$_5$ in the global phase diagram for heavy-fermion metals. The outer layer with faded color shows the global phase diagram\cite{30Si}, and the embedded is a schematic magnetic field-pressure phase diagram of CeRhIn$_5$ at $T$=0.}
\label{Fig.4}
\end{figure}

We reach the important conclusion that, upon increasing the magnetic field at ambient pressure and zero temperature, (i) an abrupt transition associated with the localization/delocalization of the 4$f$ electrons occurs at $B^{\ast}$$\approx$35T, i.e., inside the AF state of CeRhIn$_5$, and (ii) the AF QCP at $B_{c0}$$\approx$50T is of the (three-dimensional) SDW-type, consistent with the field dependence of $T_N$ as stated above. This is in contrast to what happens as a function of pressure at relatively low magnetic fields, where the dHvA oscillations indicate a 4$f$-localized/delocalized transition at the AF QCP\cite{7Shishido}. Our results, therefore, provide the direct Fermi-surface characterization of different types of QCPs in a single material accessed by tuning of different control parameters.

We interpret the two types of AF QCPs in the pressure-field phase diagram [Fig. 4(b)] in terms of the global phase diagram of a model of quantum criticality in which fermionic degrees of freedom become critical\cite{5Si,6Coleman}. This model explicitly delineates the evolution of the zero-temperature AF transition and the Kondo-destruction in a multi-parameter phase space. The transition induced by a magnetic field at ambient pressure corresponds to trajectory I; as the field is increased, Kondo resonances are switched on at $B_0^{\ast}$ before the AF order is suppressed at $B_{c0}$, and the Fermi surface evolves smoothly across the latter transition. This is in contrast to the transition caused by pressure at relatively low field (trajectory II)\cite{7Shishido}, at which the Kondo destruction and the magnetic transition take place simultaneously, leading to a jump of the Fermi surface at the zero-temperature continuous AF phase transition. We note that an alternative theoretical approach \cite{HK}, which treats the dynamical effects of the Kondo coupling but not those of the competing RKKY interaction, leads to a phase diagram that contains the type I transition but misses the type II transition.

These results on CeRhIn$_5$ shed new light on the quantum phases and their transitions in heavy-fermion systems in general. They suggest that the same type of continuous quantum phase transition inside the ordered AF region may exist in other heavy-fermion systems, such as Co-doped YbRh$_2$Si$_2$\cite{25Friedemann} and Yb$_2$Pt$_2$Pb\cite{26Kim}, as well as CeCu$_{6-x}$Au$_x$\cite{27Stockert} and Ce$_3$Pd$_{20}$Si$_6$\cite{28Custers} when tuned by a magnetic field. In CeIn$_3$, dHvA frequencies and the corresponding cyclotron masses of heavy-hole pockets undergo a sharp increase near 40T, which is below the critical field\cite{29Sebastian}. This suggests the possibility of a similar localization/delocalization transition inside its AF phase, even though dHvA oscillations associated with the large Fermi surface have not yet been observed at high fields.

The evolution of the Fermi surface as a function of magnetic field in CeRhIn$_5$ demonstrates the robustness of an electronic reconfiguration as inferred previously from the dHvA results as a function of pressure\cite{7Shishido}. Furthermore, it allows us to place the zero-field pressure-induced superconductivity in the theoretically proposed global phase diagram\cite{5Si,6Coleman}. In contrast to the application of a high field at ambient pressure, the low-field region under pressure is located already in the quantum critical portion of the phase diagram, and so should be the zero-field region under pressure (Fig. 4(b)). That is, the putative AF QCP inside the pressure-induced superconducting dome is likely to be of the unconventional type. This suggests that heavy-fermion superconductivity not only arises in the vicinity of SDW-type QCPs\cite{11Arndt}, but can also be driven by electronic fluctuations arising from an unconventional QCP.

Our findings point to two important lessons in the broader context of quantum materials. First, because CeRhIn$_5$ has one of the highest superconducting transition temperatures among 4$f$-based heavy-fermion metals, our results indicate that the fragile electronic excitations associated with the localization-delocalization transition of the 4$f$ states can promote robust superconductivity. This provides a link between the superconductivity of 4$f$-electron-based heavy fermions and that of other strongly correlated electron systems, including the doped Mott insulators of the 3$d$-electron-based copper oxides and the $sp$-electron-based organic charge-transfer salts. Second, our work demonstrates that direct measurements of the Fermi surface topology can distinguish models of quantum criticality and, as our work demonstrates, reveal multiple classes of QCPs. This step toward a unifying characterization of criticality in heavy-fermion metals anticipates a universal description of their quantum-driven phase transitions.

\acknowledgments

We acknowledge valuable discussion with R. Daou and N. Harrison. Work at Zhejiang University is supported by the National Basic Research Program of China (973 Program) (Grant Nos. 2009CB929104 and 2011CBA00103), the NSFC (Grant Nos.10934005, 11174245 and 10874146), Zhejiang Provincial Natural Science Foundation of China and the Fundamental Research Funds for the Central Universities. Work at LANL is performed under the auspices of the DOE and was supported in part by the DOE/Office of Science, NSF, State of Florida and the LANL LDRD program. TP acknowledges support from NRF (Grant No. 220-2011-1-C00014). Work at Dresden is supported by the DFG Research Unit 960 ¡°Quantum Phase Transitions¡±. Work at Rice University is in part supported by NSF (Grant No. DMR-1006985) and the Robert A. Welch Foundation (Grant No. C-1411).

\end{document}